\documentclass[12pt,preprintnumbers]{revtex4}
\usepackage{epsfig}

\def\eq#1{{(\ref{#1})}}
\def\fig#1{{Fig.~\ref{#1}}}

\newcommand{\beq}{\begin{equation}}
\newcommand{\eeq}{\end{equation}}
\newcommand{\beqar}[1]{\begin{eqnarray}\label{#1}}
\newcommand{\eeqar}{\end{eqnarray}}

\newcommand{\as}{\alpha_s}
\newcommand{\un}{\underline}

\newcommand{\stackeven}[2]{{{}_{\displaystyle{#1}}\atop\displaystyle{#2}}}
\newcommand{\lsim}{\stackeven{<}{\sim}}

\begin{document}

\preprint{BNL-NT-04/2}

\title {
 Heavy quark production by a quasi-classical color field
in proton-nucleus collisions 
}

\author{Kirill Tuchin}

\affiliation{ Physics Department, Brookhaven National Laboratory,\\
Upton, NY 11973-5000, USA}

\date{\today}

\begin{abstract} 

We calculate the inclusive heavy quark production cross section for 
proton-nucleus collisions at high energies. We perform calculation 
in a quasi-classical approximation (McLerran-Venugopalan model) 
neglecting all low-$x$ evolution effects. The derived expression for the 
differential cross section  can be applied for studying the heavy quark 
production in the central rapidity region at RHIC.

\end{abstract}
\maketitle 


In this letter we address a problem of heavy quark production in 
proton-nucleus collisions at very high energy. 
Heavy quarks are a very important tool for studying the
properties of the strong interactions. At not very high energies 
the heavy quark  mass provides a scale which allows one to use the 
perturbative QCD \cite{collcharm} since the long distance dynamics 
is effectively  decoupled \cite{Appelquist:tg}. 
However, at high energies which correspond to low values of Bjorken $x$ 
a nucleus becomes a strongly coupled dense partonic system (Color Glass 
Condensate)  with large typical transverse momentum $Q_s$ determined by 
density of nuclear partons over the nucleus transverse area 
\cite{GLR,Mueller:wy,Blaizot:nc,MV}. Experimental data suggest
that $Q_s^2\simeq 2$~GeV$^2$ for the Gold nucleus at $x\simeq 
0.01$\cite{KL}. It is existence of the strong color field 
which violates the decoupling of the heavy quark production
subprocess from the dynamics of partons in the nucleus wave
function \cite{LRSS,KhT}.
 
The Color Glass Condensate starts to play a significant role in 
scattering processes at low-$x$ since a coherence length of gluons in a 
nuclear wave function is of the order $1/(x M_N)$ \cite{Ioffe,FS} which 
allows them to 
coherently interact with all nucleons in a nucleus. It was argued in 
refs.~\cite{MV,jmwlk0,YK} that the color field of a nucleus in a 
low-$x$ regime is 
given by the classical solution to the Yang-Mills equations. It includes 
all multiple rescatterings of a gluons with the color charges of a 
nucleus \cite{YK,jmwlk0}. However, the quasi-classical approach is not 
sufficient when $x<e^{-1/\as}$. In that case quantum evolution effects 
become important and must be resumed using the low-$x$ evolution 
equations \cite{BFKL,GLR,BK,jmwlk0,jmwlk}. 

In this paper we undertake the first step towards solution of a 
problem of heavy quark production in pA collisions at high 
energies by deriving  the heavy quark production differential 
cross section \eq{main} in a quasi-classical approximation which is 
equivalent to inclusion of all multiple rescatterings of a proton with a 
nucleus.

The process of  heavy quark production at high energies in a
quasi-classical approximation has three separated in time stages in the
nucleus rest frame. Emission of a gluon $g$ by a proton's
valence quark $q_v$ takes much longer time $\tau_{q_v\rightarrow q_vg}$ 
than a
subsequent emission of a $q\bar q$ pair by a gluon $ t_{g\rightarrow 
q\bar q} \ll \tau_{q_v\rightarrow q_vg}$. In turn, the time of 
interaction of a $q_vgq\bar q$ system  with 
a nucleus is of the order of nuclear length $R_A$ and is negligible as
compared to the evolution time of the proton wave function
$\tau_{q_v\rightarrow q_vg}\gg t_{g\rightarrow  q\bar q}\gg R_A$. 
Indeed, assume that 
proton is moving in the "+'' light cone direction with four momentum 
$p=(p^+,0,\un 0)$ and nucleus is at rest. Denote the emitted gluon 
four-momentum by 
$q=(\varsigma p^+, \frac{\un p^2}{\varsigma p^+}, \un p)$. By 
uncertainty 
principle emission of a gluon by a valence quark takes time
\beq
\tau_{q_v\rightarrow qg}= \frac{2}{q^++q^-+(p-q)^++(p-q)^--p^+-p^-}
=\frac{2\varsigma(1-\varsigma)  p_+}{\un q^2}\approx \frac{2 q^+}{\un 
q^2}
\eeq
since the emission of a gluon at high energy is dominated by 
$\varsigma\ll 1$. The Bjorken $x$ is defined as $x=\frac{\un 
q^2}{2q^+ M_N}$, where $M_N$ is a nucleon mass. Therefore
\beq\label{lifetime}
\tau_{q_v\rightarrow qg}=\frac{1}{x\, M_N}\gg R_A\simeq 
\tau_\mathrm{int}
\eeq
for very low $x$.
The same argument applies to the successive emission of a quark-antiquark 
pair by a gluon in a proton's wave function: 
a $q_v\rightarrow q_vg$ fluctuation spans much longer time 
than $g\rightarrow q\bar q$. Therefore, processes in which gluon or 
heavy quarks are produced in course of the rescatterings in a nucleus are 
suppressed by powers of energy $p^+$. 

\begin{figure}
\begin{center}
\epsfig{file=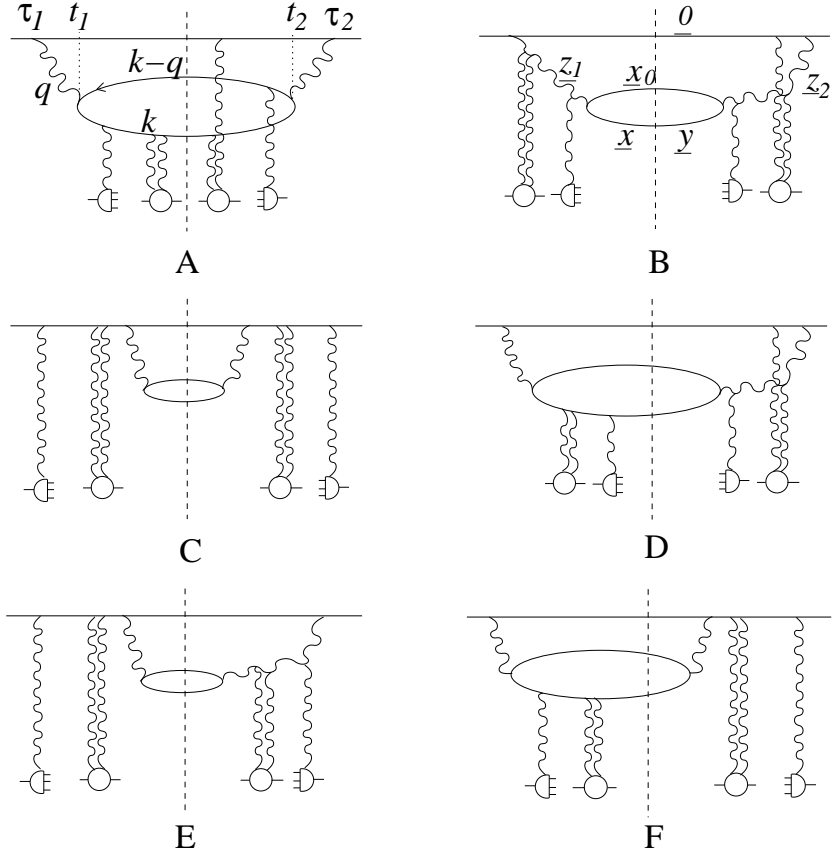,width=11cm}
\caption{ Diagrams which contribute to the heavy quark 
production in the light-cone perturbation theory. A: $\tau_1<0$, 
$t_1<0$, 
$\tau_2<0$, $t_2<0$,
B: $\tau_1<0$, $\tau_2<0$, $t_1>0$, $t_2>0$,  C: $\tau_1>0$, 
$t_1>0$, 
$\tau_2>0$, $t_2>0$, D: $\tau_1<0$,  $t_1<0$, $\tau_2<0$, $t_2>0$, 
E:  $\tau_1>0$, 
$t_1>0$,  $\tau_2<0$, $t_2>0$,   
F:  $\tau_1<0$, $t_1<0$, $\tau_2>0$, $t_2>0$.
Not shown are the complex conjugates D$^*$: $\tau_1<0$, $t_1>0$,  
$\tau_2<0$, $t_2<0$, E$^*$:  $\tau_1<0$, $t_1>0$,  $\tau_2>0$, $t_2>0$,
 F$^*$: 
$\tau_1>0$, $t_1>0$,  $\tau_2<0$, $t_2<0$. Instantaneous interaction of 
a $q_vgq\bar q$ system with the nucleus happens at light-cone time 
$\tau_\mathrm{int}=0$. The final state is denoted by the vertical dashed 
line at $\tau=\infty$.  \label{nine}}
\end{center}
\end{figure}

Let us choose the $A_+=0$ light cone gauge. 
In view of the above argument we can separate 
the process of heavy quark production in nine terms according to the 
time of gluon emission in the amplitude $\tau_1$ and in the complex 
conjugate one $\tau_2$, time of quark-antiquark emission in the 
amplitude $t_1$ and in the complex 
conjugate one  $t_2$, and the time of interaction which happens at light-cone 
time $\tau_\mathrm{int}=x_+=0$. In \fig{nine} we show all possible 
cases.  

To proceed we need to know the light-cone wave functions of a
valence quark  and of a virtual gluon in transverse configuration 
space.  The light-cone wave function of a valence quark in momentum 
space is given by 
\beq\label{vquarkm}
\psi_{q_v\rightarrow q_v g}(\un q)=g\, T^a\, 
\frac{\un\epsilon^\lambda\cdot \un q }{\un q^2},
\eeq
where $\un q$ is the gluon's transverse momentum and $\epsilon^\lambda$ 
is the gluon's polarization vector. Its Fourier image reads 
\beq\label{vquarkc}
\psi_{q_v\rightarrow q_v g}(\un z)=\int\,\frac{d^2q}{(2\pi)^2}\,
e^{-i\,\un q \cdot \un z}\, \psi_{q_v\rightarrow q_v g}(\un q)=
g\, T^a\,\frac{1}{2\pi i}\, \frac{\un\epsilon^\lambda\cdot \un z }{\un 
z^2},
\eeq  
Averaging square of Eq.~\eq{vquarkc} over quantum numbers of the 
initial quark and summing over quantum numbers of the final quark 
and gluon we obtain the familiar gluon radiation kernel of a dipole 
model \cite{dipole} 
\beq\label{lcv}
\Phi_{q_v\rightarrow q_v g}(\un z_1, \un z_2)=
\frac{1}{2N_c}\,\sum_{a,\lambda}
\psi_{q_v\rightarrow q_v g}(\un z_1)\psi^*_{q_v\rightarrow q_v g}(\un 
z_2)=
\frac{\as C_F}{2\pi}\,
\frac{\un z_1\cdot \un z_2}{\un z_1^2\,\un z_2^2},
\eeq
where $\un z_1$ and $\un z_2$ are the transverse coordinates of the 
gluon in the amplitude and in the complex conjugated amplitude 
correspondingly.

Light-cone wave function of a virtual gluon of momentum 
$q$ reads,  see \fig{nine}
\beq\label{vgluonm}
\psi_{g\rightarrow q\bar q}(\un k, \un k-\un q,\alpha)=
\frac{g\,T^a}{(\un k-\alpha\, \un q)^2+m^2}\,(\delta_{r,r'}(\un k-
\alpha \,\un q)\cdot \un 
\epsilon^\lambda\,[r(1-2\alpha)+\lambda]+r\,\delta_{r,-r'}\,m\,(1+r\lambda)),
\eeq
where $\un k$ is the produced quark's transverse momentum, $m$ its mass, 
$\alpha=k^+/q^+$ is the 
fraction of the gluon's light-cone momentum it carries, 
$r$ and $r'$ are the quark and the antiquark helicities 
correspondingly. Eq.~\eq{vgluonm} can 
be written in transverse configuration space using modified Bessel 
functions 
\begin{eqnarray}\label{vgluonc}
&&\psi_{g\rightarrow q\bar q}(\un z_1,\un x, \un x_0,\alpha)=
\int\frac{d^2k}{(2\pi)^2}\,e^{-i\,\un k \cdot (\un x-\un x_0)}\, 
\int\frac{d^2q}{(2\pi)^2}\,e^{-i\,\un q \cdot (\un x_0-\un z_1)}\,
\psi_{g\rightarrow q\bar q}(\un k, \un k-\un q,\alpha)\nonumber\\
&&=\delta((\un x_0-\un z_1)+\alpha\,(\un x-\un x_0))\,\frac{g 
T^a}{2\pi}\,
\bigg(i\,\delta_{r,r'}\,\frac{(\un x-\un x_0)\cdot \un 
\epsilon^\lambda}{|\un x-\un x_0|}\,m\, K_1(|\un x-\un 
x_0|\,m)\,[r(1-2\alpha)+\lambda]  \nonumber\\
&& 
+K_0(|\un x-\un x_0|\,m)\, r\, \delta_{r,-r'}\, m\, 
(1+r\lambda)\bigg),
\end{eqnarray}
where $\un x_1$ and $\un x_0$ are  the quark's and antiquark's 
transverse 
coordinates in the amplitude correspondingly, see \fig{nine} and
$x\equiv|\un x|$. Averaging square of Eq.~\eq{vgluonc} over 
 quantum numbers of the
initial gluon and summing over quantum numbers of the final quark 
and antiquark \cite{KM} we find 
\begin{eqnarray}\label{lcg}
\Phi_{g\rightarrow q\bar q}(\un z,\un x, \un x_0, \alpha)&=&
\frac{\as}{\pi}\, m^2\,\bigg(\, \frac{(\un x-\un x_0)\cdot   
(\un y -\un x_0)}{|\un x-\un x_0|\, |\un y -\un x_0|}
K_1(|\un x-\un x_0|\,m)\,K_1(|\un y -\un x_0|\,m)\nonumber\\
&&\times[\,\alpha^2\,+\,(1-\alpha)^2\,]
+\,K_0(|\un x-\un x_0|\,m)K_0(|\un y -\un x_0|\,m) \,\bigg),
\end{eqnarray}
where $\un y$ is the quark's transverse coordinate in the complex 
conjugate amplitude and we do not include two delta functions (see 
\eq{vgluonc}) in 
definition of $\Phi_{g\rightarrow q\bar q}$. Eq.~\eq{lcg} is a special 
case of light-cone 
wave function of an off-shell gauge boson derived in Refs.~\cite{KM,NZ}. 

\begin{figure}
\begin{center}
\epsfig{file=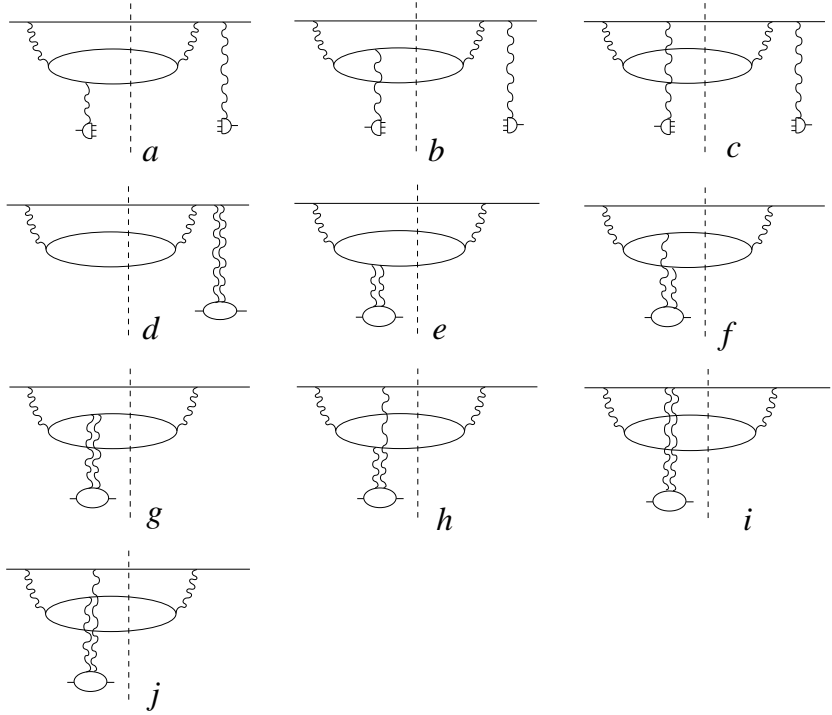,width=11cm}
\caption{ Diagrams contributing to the time ordering of the 
diagram \emph{F} in \fig{nine}.}
\label{figf}
\end{center}
\end{figure}

Rescatterings of the produced partonic system in a nucleus must be 
calculated separately for each time ordering \cite {KoM} as shown in 
\fig{nine}. 
The result can be written in terms of the Fourier 
transformation of the normalized gluon-nucleon cross section 
\cite{BDMPS,KoM}:
\beq\label{V}
V(\un x)=\int d^2l\, e^{-i\un l\cdot \un x}
 \frac{1}{\sigma}\frac{d\sigma}{d^2 l}.
\eeq 
All diagrams contributing to the time ordering of the diagram \emph{F} 
in \fig{nine} are shown  in \fig{figf}.
The sum of diagrams \emph{a-j} in \fig{figf} is given by
\begin{eqnarray}\label{sum2b}
F&=&\frac{1}{2N_c}\left(N_c-\frac{2}{N_c}\right)V(\un 
x) + \frac{1}{N_c^2}V(\un x_0)-\frac{1}{2N_c^2}V( 0)-
\frac{C_F}{2N_c}V(0)\nonumber\\
&& - \frac{C_F}{2N_c}V(0) - \frac{1}{2N_c^2}V(\un x -\un x_0)-
\frac{C_F}{2N_c}V(0)+\frac{1}{N_c^2}V( 0)\nonumber\\
&& -\frac{C_F}{2N_c}V(\un 
0)+\frac{1}{2N_c}\left(N_c-\frac{2}{N_c}\right)V(\un
x_0)\nonumber\\
&=&\frac{1}{2}\left(V(\un x)-V(0)\right)+\frac{1}{2}\left(V(\un 
x_0)-V(0)\right)-\frac{1}{2N_c^2}\left(V(\un x -\un 
x_0)-V(0)\right),
\end{eqnarray}
where $\un x$ and $\un y$ are coordinates of the produced quark in 
the amplitude 
and in the complex conjugate one correspondingly, $\un x_0$ is 
coordinate 
of antiquark, see \fig{nine}. Multiplying expression \eq{sum2b} by the 
nucleus 
profile 
function $T(\un b)$, nuclear density $\rho$ and the gluon-nucleon 
cross section $\sigma$ \cite{KoM} we obtain
\beq\label{p1}
-P(\un x, \un x_0)= -\frac{1}{8}\,\un x^2 \, Q_s^2-
\frac{1}{8}\,\un x_0^2 \, Q_s^2+\frac{1}{8N_c^2}\, (\un x-\un 
x_0)^2\, Q_s^2,
\eeq
where we follow notations of \cite{YKdif}.
The saturation scale $Q_s^2$ in \eq{p1} is given by \cite{dipole,KoM}
\beq\label{satscale}
Q_s^2(\un x)=\frac{4\pi^2\as N_c}{N_c^2-1}\,\rho\, T(\un b)\,
xG(x,1/\un x^2),
\eeq 
where the gluon distribution function in a nucleon reads
\beq\label{xG}
xG(x,1/\un x^2)=\frac{\as\, C_F}{\pi}\,\ln\frac{1}{\un x^2\mu^2},
\eeq
with $\mu$ some infrared cutoff. For spherical nucleus $T(\un b)=
2\sqrt{R^2-\un b^2}$.
Assuming that the interactions of a proton with individual nucleons are 
independent \cite{dipole} we can exponentiate the formula \eq{p1} to 
obtain for the 
diagram F on \fig{nine} 
\beq\label{D}
F=\exp\left\{-P(\un x, \un x_0)\right\}=
\exp\left\{ -\frac{1}{8}\,\un x^2\, Q_s^2-
\frac{1}{8}\,\un x_0^2 Q_s^2+\frac{1}{8N_c^2}\,(\un x-\un x_0)^2\,Q_s^2 
\right\}.
\eeq
This formula coincides with the $q\bar qg$ ``propagator" derived in 
refs.~\cite{Kopeliovich,KWi,YKdif}.

\begin{figure}
\begin{center}
\epsfig{file=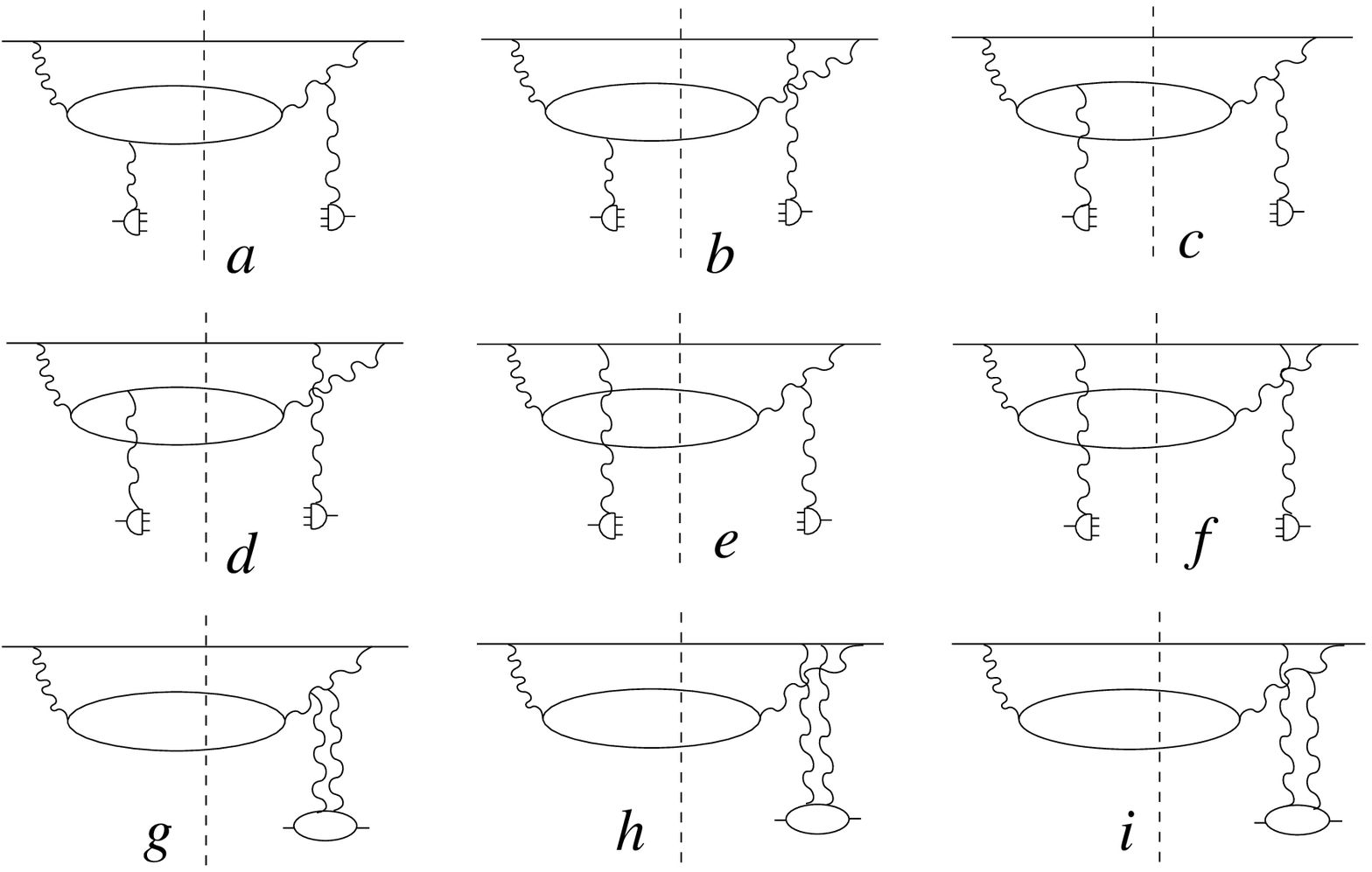,width=11cm}
\caption{ Diagrams contributing to the time ordering of the
diagram D in \fig{nine}}\label{figd}
\end{center}
\end{figure}

Analogously, all diagrams contributing to the time ordering of the 
diagram D in \fig{nine} are shown  in \fig{figd} and in 
\fig{figf}: \emph{e-j}.
The sum of diagrams \emph{a-i} in \fig{figd} and \emph{e-j} in 
\fig{figf} yields
\begin{eqnarray}\label{sum2d}
D&=& \frac{1}{2}V(\un x-\un z_2)-\frac{1}{N_c^2}V(\un 
x)+\frac{1}{2}
V(\un x_0-z_2)- \frac{1}{2N_c}\left(N_c-\frac{2}{N_c}\right)V(\un
x_0)\nonumber\\
&& -\frac{1}{2}V(\un z_2)+\frac{C_F}{N_c}V(0)-\frac{1}{2}V(0)-
\frac{C_F}{2N_c}V(0)\nonumber\\
&& +\frac{1}{2}V(\un z_2)
 - \frac{C_F}{2N_c}V(0) - \frac{1}{2N_c^2}V(\un x -\un x_0)-
\frac{C_F}{2N_c}V(0)\nonumber\\
&&+\frac{1}{N_c^2}V( 0) -\frac{C_F}{2N_c}V(0)-
\frac{1}{2N_c}\left(N_c-\frac{2}{N_c}\right)V(\un
x_0)\nonumber\\
&=&\frac{1}{2}\left(V(\un x -\un z_2)-V(0)\right)+\frac{1}{2}\left(V(\un
x_0-\un z_2)-V(0)\right)-\frac{1}{2N_c^2}\left(V(\un x -\un
x_0)-V(0)\right).
\end{eqnarray}
Multiplying \eq{sum2d} by $T(\un b)\, \rho\, \sigma$ and exponentiating
we derive
\beq\label{F}
D=\exp\left\{-P(\un x-\un z_2, \un x_0-\un z_2)\right\}
\eeq
where we used definition \eq{D}. Complex conjugated of F and D can be 
obtained by replacing 
$\un x\leftrightarrow \un y$, and $\un z_1 \leftrightarrow \un z_2$:
\begin{eqnarray}
D^*&=&\exp\left\{-P(\un y, \un x_0)\right\},\label{D*}\\ 
F^*&=&\exp\left\{-P(\un y-\un z_1, \un x_0-\un z_1)\right\}. \label{F*}
\end{eqnarray}
Diagrams \emph{B},\emph{C} and \emph{E}, \emph{E$^*$}
have been calculated in \cite{KoM}. The only difference is 
additional color factor $\frac{1}{2}$ 
emerging due to fluctuation of a virtual gluon into quark--anti-quark 
pair. We included this factor in the definition of the wave function 
\eq{lcg}. We have 
\begin{eqnarray}
B&=& e^{-\frac{1}{4}\,(\un z_1-\un z_2)^2\,Q_s^2}\label{B}\\
E&=& e^{-\frac{1}{4}\,\un z_2^2\,Q_s^2}\label{E}\\
E^*&=& e^{-\frac{1}{4}\,\un z_1^2\,Q_s^2}\label{E*}
\end{eqnarray}
Finally, it is easy to see that the diagram A in \fig{nine} is equal to
\beq\label{A}
A=e^{-\frac{1}{4}\,\frac{C_F}{N_c}\,(\un x-\un y)^2\, Q_s^2}.
\eeq
Summing up diagrams  A -- E and their complex conjugates
results in the following rescatterings factor
\begin{eqnarray}\label{xi}
\Xi(\un x, \un y, \un x_0, \un z_1, \un z_2)&=&
e^{-\frac{1}{4}\, (\un z_1-\un z_2)^2\, Q_s^2}- e^{-\frac{1}{4}\, \un 
z_1^2\, Q_s^2}
-e^{-\frac{1}{4}\, \un z_2^2\, Q_s^2}+
e^{-\frac{1}{4}\, \frac{C_F}{N_c}\, (\un x-\un y)^2\, Q_s^2}
\nonumber\\
&& + e^{-P(\un x, \un x_0)}+e^{-P(\un y, \un x_0)}-
e^{-P(\un x-\un z_2, \un x_0-\un z_2)}-e^{-P(\un y-\un z_1, \un 
x_0-\un z_1)}
\end{eqnarray}

Using \eq{lcv}, \eq{lcg} and \eq{xi} we can write 
down  the inclusive quark production cross section
\begin{eqnarray}\label{main}
&&\frac{d\sigma}{d^2k \,dy}=
\int d^2b\, d^2z_1\, d^2z_2\,\frac{\as\, C_F}{\pi^2}\,
\frac{\un z_1\cdot \un z_2}{\un z_1^2\,\un z_2^2}\,
\int d^2x_0
\int d\alpha\,
\int \frac{d^2x\, d^2 y}{(2\pi)^3}\, \Phi_\mathrm{g\rightarrow 
q\bar q}
(\un x-\un x_0, \un y-\un x_0, \alpha)\, 
\nonumber\\ 
&&
\times
e^{-i\un k\cdot (\un x-\un y)}\,
\Xi(\un x, \un y, \un x_0, \un z_1, \un z_2)\,
\delta((\un x_0-\un z_1)+\alpha\,(\un x-\un x_0))\,
\delta((\un x_0-\un z_2)+\alpha\,(\un y-\un x_0)).
\end{eqnarray}
where $\un b$ is an impact parameter. This formula is a generalization of  
result obtained by Kopeliovich and Tarasov in Ref.~\cite{KopTar}.

Formula \eq{main} is the main result of our paper.  It resums all higher 
twist effects in the quasi-classical approximation which means that we 
keep all terms proportional to $\as^2 A^{1/3}\sim 1$ and neglect terms 
suppressed by powers of $\as\ll 1$. We explicitly neglected the low-$x$ 
quantum evolution effects assuming that $\as\ln(1/x)\ll 1$. Therefore 
formula \eq{main} is applicable when $e^{-1/\as}\lsim x\ll 1$. 
As resent experimental data on $dA$ collisions at RHIC show, this 
corresponds to the central rapidity region  at $\sqrt{s}=200$~GeV 
\cite{Debbe}. 

Formula \eq{main} has been used in Ref.~\cite{KhT} for numerical 
calculations of charm production at RHIC. It was shown that the 
charm spectrum obtained according to \eq{main} is much harder than in 
naive parton model approach. This is attributed to the presence of a hard 
`intrinsic' scale $Q_s^2$. It is clear that the dependence of a heavy 
quark yield on $A$ is closely related to the relation between $Q_s$ and 
$m$. In the strong color field $Q_s\gg m$ the total cross section  
of heavy quark production in pA collisions is proportional to the 
transverse size of a nucleus 
$\sigma_\mathrm{tot}\sim A^{2/3}$ due to saturation in a nucleus wave 
function. 
In the opposite limit $Q_s\ll m$ the color field of a nucleus is not able 
to produce heavy quarks from the vacuum in which case 
$\sigma_\mathrm{tot}\sim A$. Therefore, at high energies one expects 
suppression of the heavy quark yield. In the case of charm quark 
production numerical calculations in \cite{KhT} show that at $y=0$ at 
RHIC the charm quark yield is not suppressed.  However at forward 
rapidities it gets suppressed since the nuclear color  field strength 
increases at small $x$ due to quantum evolution. I refer 
the reader interested in phenomenological applications of \eq{main} to 
Ref.~\cite{KhT} for more detailed discussion. 

Dynamics of saturated quasi-classical color fields dominates the total 
multiplicities of AA and dA collisions in the central rapidity region at 
RHIC \cite{KL,KN,KLN,KLd,KV}. The 
Cronin enhancement seen in the data \cite{dAdata} is produced by 
multiple 
rescatterings of a proton \cite{GW,KNST} in a saturated wave function of a 
nucleus \cite{cronin,JNV,KW}.  These multiple rescatterings produce
particle correlations which give a substantial contribution to the
elliptic flow phenomenon in AA collisions \cite{flow}. 
Eq.~\eq{main} can be used for analysis of the
heavy quark production in pA collisions at the central rapidity region at
RHIC. In particular, one can address the question of whether formula 
\eq{main} yields the Cronin enhancement of charm production 
analogously to the case of gluon production \cite{cronin,KW,JNV}. 

High energy quantum evolution has been neglected throughout this paper. 
However, as energy/rapidity increases the quantum evolution becomes 
important \cite{jmwlk0,jmwlk,BK,LT}. It gives rise 
to a number of  spectacular effects \cite{KLM,KhT,cronin,KW} associated 
with the extended geometric scaling phenomenon \cite{geom,LT2,IIM,MT}. 
Recent results of BRAHMS collaboration \cite{Debbe} at RHIC indicate 
onset of the 
high energy evolution at rapidities close to  the proton fragmentation 
region in agreement with theoretical predictions. Therefore, 
generalization of \eq{main} to include low-$x$
quantum evolution is an important task which will be pursued in our 
forthcoming publications.

Finally, we would like to mention an important theoretical 
question which have not been touched in this paper. It is whether 
eq.~\eq{main} can be reduced to the $k_T$-factorized form. The 
$k_T$-factorization was proved for \emph{dilute} target regime in 
\cite{LRSS,CCH,CE} and have  been recently rederived in a Color Glass 
Condensate framework in 
ref~\cite{GV}. It was also proved in ref.~\cite{KTin} that the inclusive 
gluon production cross section in pA collisions can be reduced to the 
$k_T$-factorized form even if the quantum evolution is included. So far 
all phenomenological studies of the heavy quark hadroproduction at high 
partonic densities \cite{apple,KhT}
have been based on $k_T$-factorization. Therefore  it is important to 
understand to what extend it can be realized at high energies and/or 
for heavy nuclei.
We are going to address this problem elsewhere.


\vskip0.3cm
{\large\bf Acknowledgments}

The author is  indebted to Yuri Kovchegov for a continuous 
fruitful discussions 
of saturation physics over many years. 
The author gratefully acknowledges stimulating and helpful discussions  
with 
Francois Gelis, Dmitri Kharzeev, Eugene Levin, Larry McLerran and Raju 
Venugopalan.
This research was supported by the U.S. Department of
Energy under Grant No. DE-AC02-98CH10886.


\end{document}